\documentclass[aps,showpacs,floatfix,twocolumn,amsmath,amssymb]{revtex4}
\usepackage{epsfig} 

\begin{document}

\title{Magnetic moments of vector, axial, and tensor mesons in lattice QCD}

\author{Frank X. Lee and Scott Moerschbacher}
\affiliation{Physics Department,
The George Washington University, Washington, DC 20052, USA}
\author{Walter Wilcox}
\affiliation{Department of Physics, Baylor University, Waco, TX 76798, USA}

\begin{abstract}
We present a calculation of magnetic moments for selected 
spin-1 mesons using the techniques of lattice QCD. 
This is carried out by introducing progressively small static magnetic field 
on the lattice and measuring the linear response of a hadron's mass shift. 
The calculations are done on $24^4$ quenched lattices using standard Wilson actions,
with $\beta$=6.0 and pion mass down to 500 MeV.
The results are compared to those from the form factor method.
\end{abstract}
\vspace{1cm}  
\maketitle

 \section{Introduction}
Magnetic moment is a fundamental property of hadrons that arises from  
the linear response of a bound system to an external stimulus 
(in this case, magnetic field). 
It is a good testing ground for studying the internal structure of hadrons 
as governed by the quark-gluon dynamics of QCD, the fundamental theory of the 
strong interaction.
Efforts to compute the magnetic moment on the lattice come in two categories. 
One is the form factor method which involves 
three-point functions~\cite{mart89,leinweber3,wilcox92,ji02,horse03,zanotti04,cloet04}. 
The other is the background field method using only two-point 
functions (mass shifts)~\cite{mart82,bernard82,smit87,rubin95}.
The form factor method requires an extrapolation to zero momentum transfer $G_M(Q^2=0)$
due to the non-vanishing minimum discrete momentum on the lattice~\cite{wilcox02}. 
The background field method, on the other hand, accesses the magnetic moment directly 
but is limited to static properties due to the use of a static field.
Here we report a calculation of the vector meson magnetic moments in this method, 
in parallel to a recent calculation in the form factor method~\cite{Hedditch07}.
We also report results mesons in the axial and tensor sectors.
Some of the preliminary results have been reported in a conference~\cite{Lee07}.
It is an extension of our earlier work on baryon magnetic moments~\cite{Lee05} and 
electric~\cite{Joe05} and magnetic polarizabilities~\cite{Lee06} in the same method.

\section{Correlation functions}
The mass of a meson can be extracted from the time-ordered, two-point correlation
function in the QCD vacuum, projected to zero momentum 
\begin{equation}
G(t)=\sum_{\vec{x}} \langle \eta(x) \eta^\dagger(0) \rangle
\label{gt1}
\end{equation}
where $\eta$ is the interpolating field of the meson under consideration.
The general form of the interpolating field  with a simple $\bar{q_1} q_2$ quark content 
can be written as
\begin{equation}
\eta=\bar{q_1} \Gamma q_2
\end{equation}
where $\Gamma$ is a general gamma matrix that depends on the meson type.
For the spin-1 mesons considered in this work, $\Gamma=\gamma_\mu$ for vector mesons, 
$\Gamma=\gamma_5\gamma_\mu$ for axial mesons, 
and $\Gamma=\gamma_\mu \gamma_\nu$ for tensor mesons.
Here we only consider mesons with $q_1$ and $q_2$ different to avoid the complication of 
disconnected loops. 

On the quark level, Eq.~(\ref{gt1}) is evaluated by contracting out the quark pairs, 
\begin{equation}
G(t)=-\sum_{\vec{x}} \mbox{Tr} \left[ 
S_{q_1}(x,0) \gamma_0 \Gamma^\dagger \gamma_0
\gamma_5 S_{q_2}^\dagger (x,0)  \gamma_5 \Gamma \right]
\label{gt2}
\end{equation}
where $S_{q}(x,0)$ denotes the fully-interacting quark propagator. 
It is defined as Euclidean-space path integrals over gauge field $G_\mu$
\begin{equation}
S_{q}(x,0) \equiv
{\int DG_\mu  \mbox{det}(M) e^{-S_G} M^{-1} \over
 \int DG_\mu  \mbox{det}(M) e^{-S_G}  }
\label{qprop}
\end{equation}
where $S_G$ 
is the gauge action of QCD and $M=\gamma^\mu D_\mu + m_q$ its quark matrix.
On the lattice, the propagator is evaluated numerically by Monte-Carlo methods.
We use quenched approximation in this work which corresponds to setting $\mbox{det}(M)$
to a constant.

On the hadronic level, the correlation function is saturated by the complete spectrum of
intermediate states
\begin{equation}
G(t)=\sum_i w_i\,e^{-m_i t}
\label{pole2}
\end{equation}
where $m_i$ are the masses and $w_i$ are spectral weights
that are a measure of the ability of the
interpolating field to excite or annhilate the states from the QCD vacuum.
The ground state can be extracted by fitting $G(t)$ at large time.

To compute magnetic moments, we need to use polarized 
interpolating fields. For a magnetic field applied in 
the z-direction, we use 
\begin{equation}
\eta_\pm={1\over \sqrt{2}}\bar{q_1}\left(\mp\Gamma_x - i\Gamma_y \right) q_2
= {1\over \sqrt{2}}\left(\eta_x \pm i\eta_y \right).
\end{equation}
The interaction energies $E_\pm$ are extracted from the correlation functions
\begin{equation}
\langle \eta_\pm \eta_\pm^\dagger \rangle = 
{1\over 2}\left[ \langle \eta_x \eta^\dagger_x  \rangle
    \pm i \left( \langle \eta_x \eta^\dagger_y  \rangle 
                -\langle \eta_y \eta^\dagger_x  \rangle \right)
                +\langle \eta_y \eta^\dagger_y  \rangle \right].
\label{polar2}
\end{equation}
Eq.~(\ref{polar2}) implies that the polarization comes from the imaginary parts of 
the off-diagonal correlation between x and y components in the presence of the magnetic field. 
These imaginary parts are zero in the absence of the field, 
so they are responsible for the magnetic moments we observe.
We use for vector mesons $\Gamma_x=\gamma_1$ and $\Gamma_y=\gamma_2$;
for axial  mesons $\Gamma_x=\gamma_5\gamma_1$ and $\Gamma_y=\gamma_5\gamma_2$;
and for tensor mesons $\Gamma_x=\gamma_2\gamma_3$ and $\Gamma_y=\gamma_1\gamma_3$.

For each meson type, different quark combinations $q_1$ and $q_2$ 
correspond to different states.
In the case of vector mesons, they are the well-known 
$\rho^+(\bar{d}u)$, $\rho^-(\bar{u}d)$, 
$\phi(\bar{s}s)$, $K^{*+}(\bar{s}u)$, $K^{*-}(\bar{u}s$, and $K^{*0}(\bar{s}d)$. 
In the case of axial mesons, they are $a_1^+(\bar{d}u)$, $a_1^-(\bar{u}d)$, 
$K_1^{*+}(\bar{s}u)$, $K_1^{*-}(\bar{u}s)$, and $K_1^{*0}(\bar{s}d)$. 
In the case of tensor mesons, they are less well-known and 
we call them $b_1^+(\bar{d}u)$, $b_1^-(\bar{u}d)$, 
$K_t^{*+}(\bar{s}u)$, $K_t^{*-}(\bar{u}s$, and $K_t^{*0}(\bar{s}d)$. 
Counting the states with $\bar{s}s$ content (like the $\phi$ meson), 
we cover 18 states of spin-1 mesons.

 \section{Background-field Method}
For a particle of spin $s$ in uniform fields, 
\begin{equation}
E_\pm=m\pm\mu B
\end{equation}
where the upper sign means spin up and the lower sign means spin-down relative 
to the magnetic field, and 
$\mu=g {e\over 2m}s$. 
We use the following method to extract the g factors,
\begin{equation}
g=m{(E_+ - m)-(E_- - m)\over eBs}.
\label{g1}
\end{equation}

In order to place a magnetic field on the lattice, we construct an 
analogy to the continuum case. The covariant derivative of QCD is modified 
by the minimal coupling prescription 
\begin{equation}
D_\mu = \partial_\mu+gG_\mu + q A_\mu
\end{equation}
where $q$ is the charge of the fermion field and $A_\mu$ is the four-vector 
potential describing the background field. On the lattice, the gluon fields $G_\mu$ are 
introduced via link variables $U_\mu(x)=\exp{(igaG_\mu)}$. So the prescription
amounts to multiplying a U(1) phase factor $\exp(iqaA_\mu)$ to the gauge links.
Choosing $A_y = B x $, a constant magnetic field B can be introduced 
in the $z$-direction. Then the phase factor is applied to the y-links
\begin{equation}
U_y \rightarrow \exp{(iqaBx)} U_y.
\end{equation}
In our calculations, we use a linearized version for small field strengths
\begin{equation}
U_y \rightarrow (1 + i\,qaBx) U_y.
\end{equation}
The computational demand of such background-field calculations can be divided 
into three categories.
The first is a {\em fully-dynamical} calculation. For each value of the field, a new dynamical ensemble is needed that couples to u-quark (q=1/3), d-and s-quark (q=-2/3). This requires a Monte Carlo algorithm that can treat the three flavors distinctively. Quark propagators are then computed on the ensembles with matching field values. This has not been attempted.
The second can be termed as the {\em re-weighting} method in which a perturbative expansion of the action in terms of the field is performed. There has been an attempt~\cite{Engel07} to compute the neutron electric polarizability in this method. It involves the evaluation of disconnected 
diagrams.
The third is what we call {\em U(1) quenched}. No field is applied in the Monte-Carlo generation of the gauge fields, only in the valence quark propagation in the given gauge background.
In this case, any gauge ensemble can be used to compute valence quark propagators. 

We use standard Wilson actions on the $24^4$ lattice at $\beta=6.0$, 
both SU(3) and U(1) quenched, and six kappa values $\kappa$=0.1515, 0.1525, 0.1535, 
0.1540, 0.1545, 0.1555, corresponding to pion mass of about 1015, 908, 794, 732, 667, 522 MeV.
The critical value of kappa is $\kappa_c$=0.1571.
The strange pion mass is set at $\kappa$=0.1535. The source location for the quark propagators 
is (x,y,z,t)=(12,1,1,2).
We analyzed 100 configurations.
The following five dimensionless numbers 
$\eta=qBa^2$=+0.00036, -0.00072, +0.00144, -0.00288, +0.00576 give four small B fields 
(two positive, two negative) at 
$eBa^2$=-0.00108, +0.00216,  -0.00432, +0.00864 for both u and d (or s) quarks. 
These field values do not obey the quantization condition for periodicity since the values given by the condition cause too strong (too large a mass shift) for the small-field-expansion 
method to work. 
To minimize the boundary effects, we work with Dirichlet boundary conditions 
in the x-direction and large $N_x$. In addition, we place the source in the middle 
of the lattice in the x-direction so that quarks have little chance of 
propagating to the edge. We also use Dirichlet boundary conditions in the t-direction 
to maximize the number of time slices for mass extraction.
To eliminate the contamination from the even-power terms, we calculate mass shifts 
both in the field $B$ and its reverse $-B$ for each value of $B$, 
then take the difference and divide by 2.
Another benefit of repeating the calculation with the field reversed is that 
by taking the average of $\delta m (B)$ and $\delta m (-B)$ in the same dataset, 
one can eliminate the odd-powered terms in the mass shift. 
The coefficient of the leading quadratic term 
is directly related to the magnetic polarizability~\cite{Lee06}.

 \section{Results and discussion}

 \subsection{Vector mesons}
\begin{figure}
\parbox{.5\textwidth}{%
\psfig{file=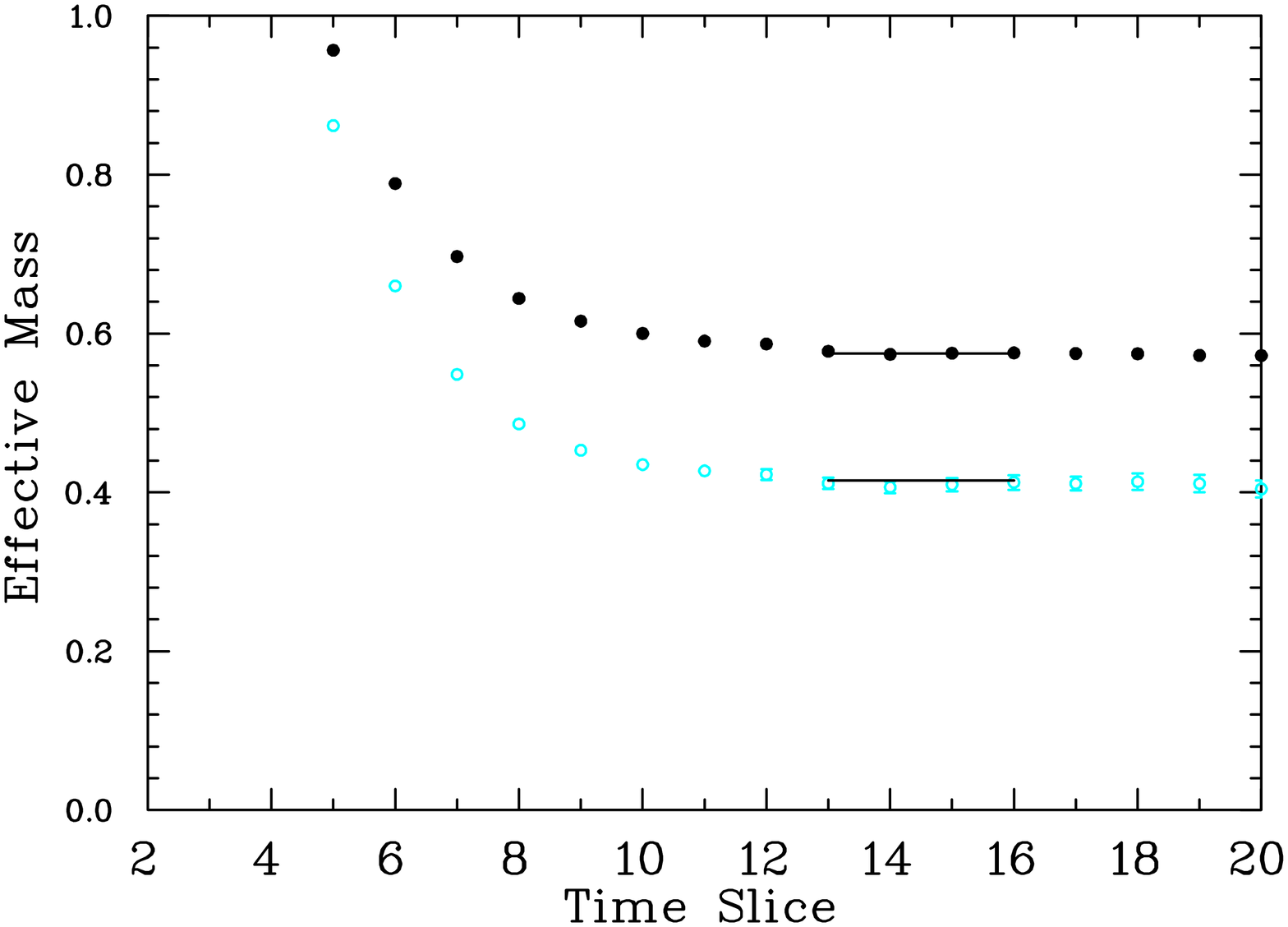,width=2.3in,angle=90}}
\parbox{.5\textwidth}{%
\psfig{file=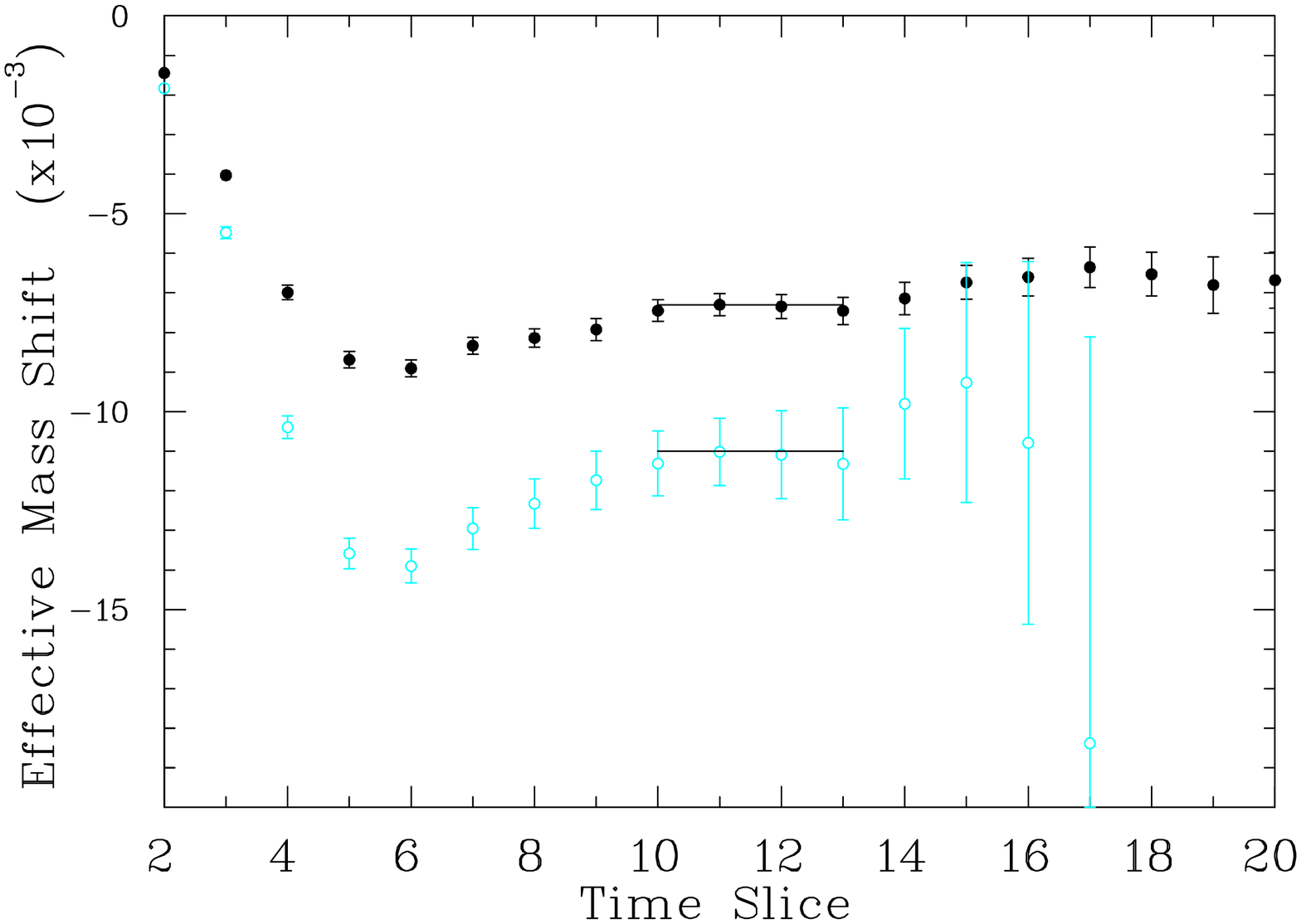,width=2.3in,angle=90}}
\caption{Effective mass plot for the $\rho^+$ vector meson mass
at zero field (top), and effective mass shifts at the weakest magnetic field (bottom)
in lattice units. The solid and empty symbols correspond to the heaviest 
and lightest pion masses, respectively.}
\label{emass2}
\end{figure}

\begin{figure*} 
\centerline{\psfig{file=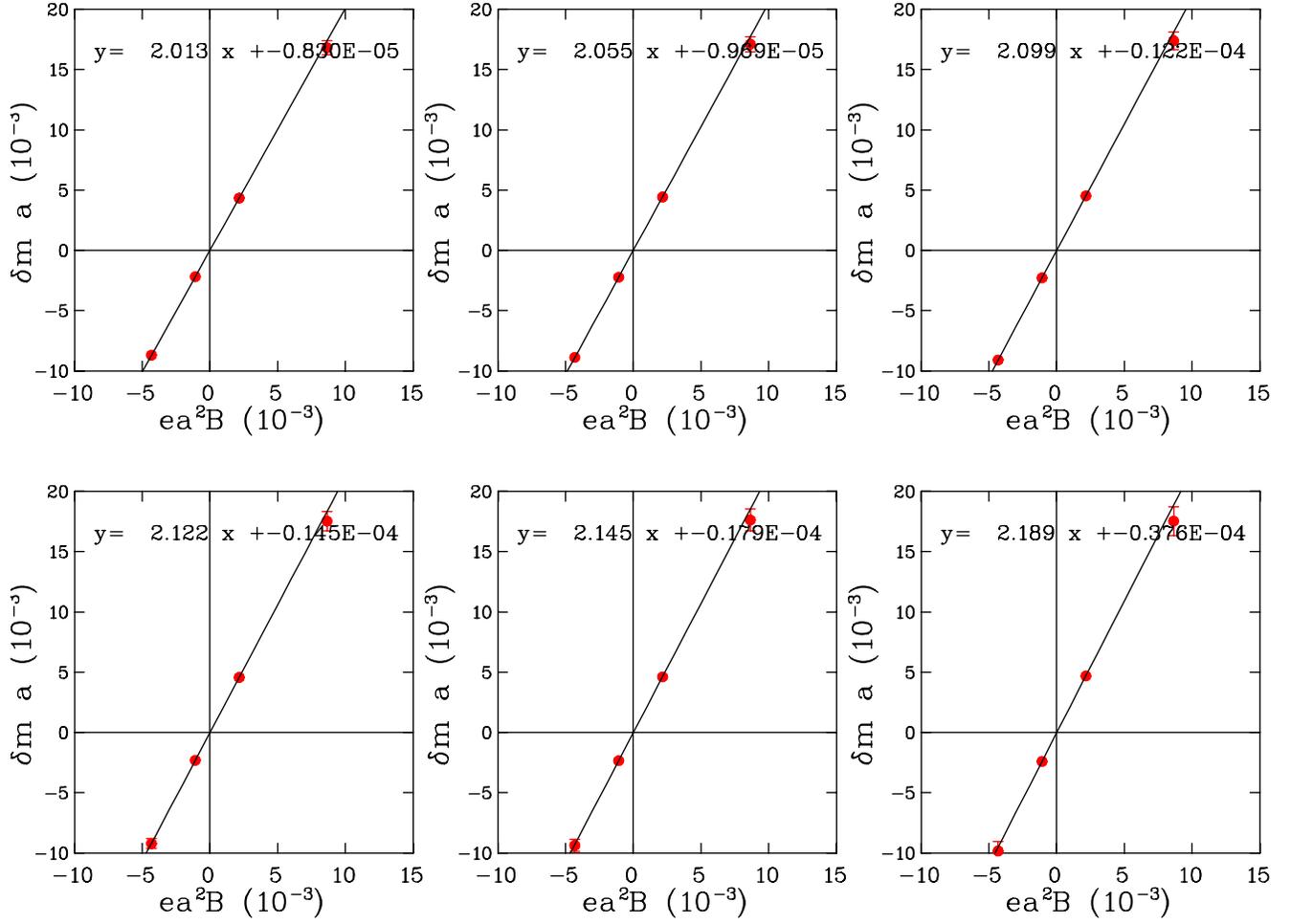,width=5.0in,angle=90}}
\caption{Mass shifts for the $\rho^+$ meson as a function of the magnetic field 
in lattice units at the six pion masses (heavy to light from top left to right, 
then to bottom left to right). 
The slope of the mass shift at each pion mass gives the g factor corresponding to 
that pion mass.
The line is a fit using only the two smallest B values.}
\label{rhop_shift_linear}
\end{figure*}

Fig.~\ref{emass2} displays a typical effective mass plot for $\rho^+$.
Both the mass and the mass shifts are shown.
Good plateaus exist for all six pion masses.
The mass shifts are extracted from the time window 10 to 13, as indicated in the figure.
Fig.~\ref{rhop_shift_linear} shows the mass shifts, 
defined as $\delta=g (eBs)$ from Eq.~(\ref{g1}), as a function of the field
for the $\rho^+$ meson. The slope gives the g-factor.
There is good linear behavior going through the origin at all the field values,
an indication that contamination from the higher-power terms has been 
effectively eliminated by the $(\delta(B)-\delta(-B))/2$ procedure. 
This is also confirmed numerically by the smallness of 
intercept as shown in the fit results $y=a x+b$. 
At the lightest pion mass, there is a slight deviation from linear behavior at the stronger fields.
For this reason, we only use the two smallest field values 
to do the linear fit at all the pion masses.

%
%

%
%

\begin{figure}
\parbox{.5\textwidth}{%
\centerline{\psfig{file=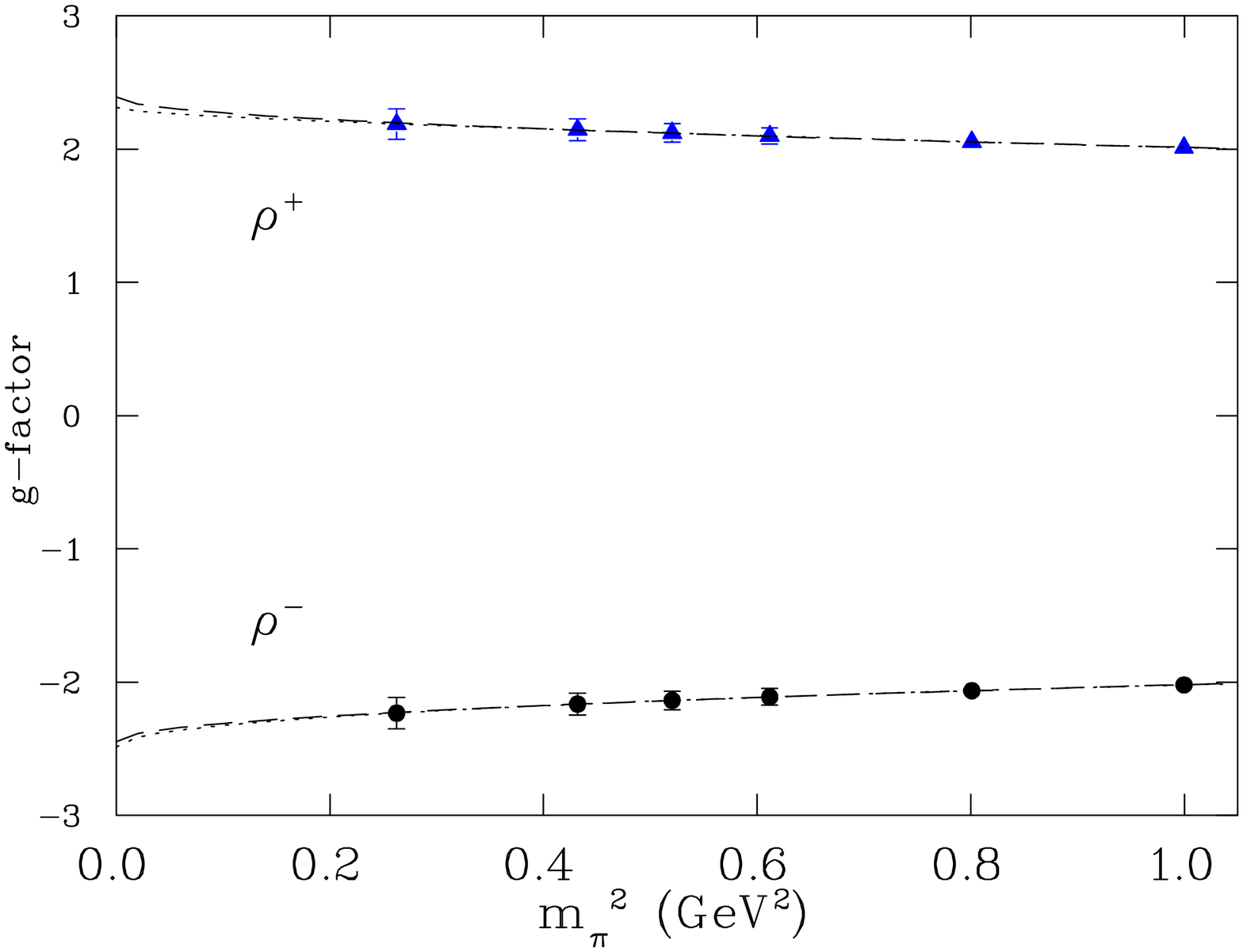,width=2.3in,angle=90}}}
\parbox{.5\textwidth}{%
\centerline{\psfig{file=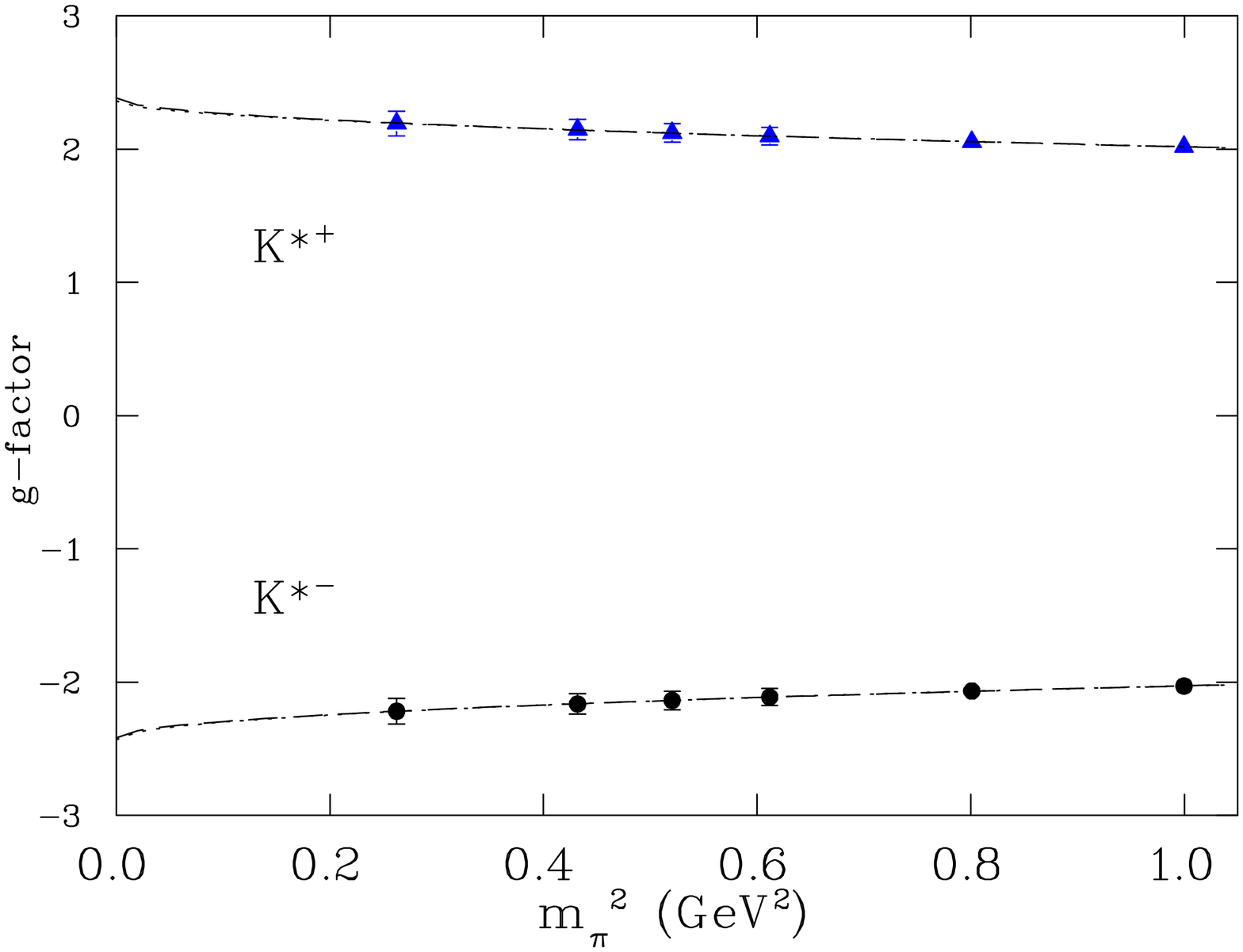,width=2.3in,angle=90}}}
\caption{G-factors for the $\rho^\pm$ (top) and $K^*$ (bottom) vector mesons 
as a function of pion mass squared. 
The 2 lines are chiral fits according to
Eq.~(\protect\ref{chiral1}) (dashed), Eq.~(\protect\ref{chiral2}) (dotted).}
\label{gfac_vector}
\end{figure}

Fig.~\ref{gfac_vector} shows the g-factors for the vector mesons as a function
of pion mass squared.
The lines are simple chiral fits using the ansatzs
\begin{equation}
g=a_0 + a_1 m_\pi,
\label{chiral1}
\end{equation}
and
\begin{equation}
g=a_0 + a_1 m_\pi+a_2 m_\pi^2.
\label{chiral2}
\end{equation}
They serve to show that there is onset of non-analytic behavior as pion mass is lowered, 
so a linear extrapolation is probably desirable.
But overall the g-factors have a fairly weak pion mass dependence. At large pion masses, 
the g-factor of $\rho^+$ approaches 2, consistent with a previous lattice calculation 
using the charge-overlap method~\cite{Wilcox97}.
Our results for $\rho^+$ are slightly higher than those from the form factor 
method (see Fig.8 in Ref.~\cite{Hedditch07}). 
The results confirmed that $g_{\rho^-}=-g_{\rho^+}$ and $g_{K^{*-}}=-g_{K^{*+}}$.
We also confirmed $g_{\rho^0}=0$ numerically (not shown). 
These relations are expected from symmetries in the correlation functions 
(these particles are charge eigenstates).
The results also show that as far as g-factors are concerned 
the $\rho$ mesons are quite similar to their strange counterparts ${K^{*}}$ mesons.

\begin{figure}[!htp]
\centerline{\psfig{file=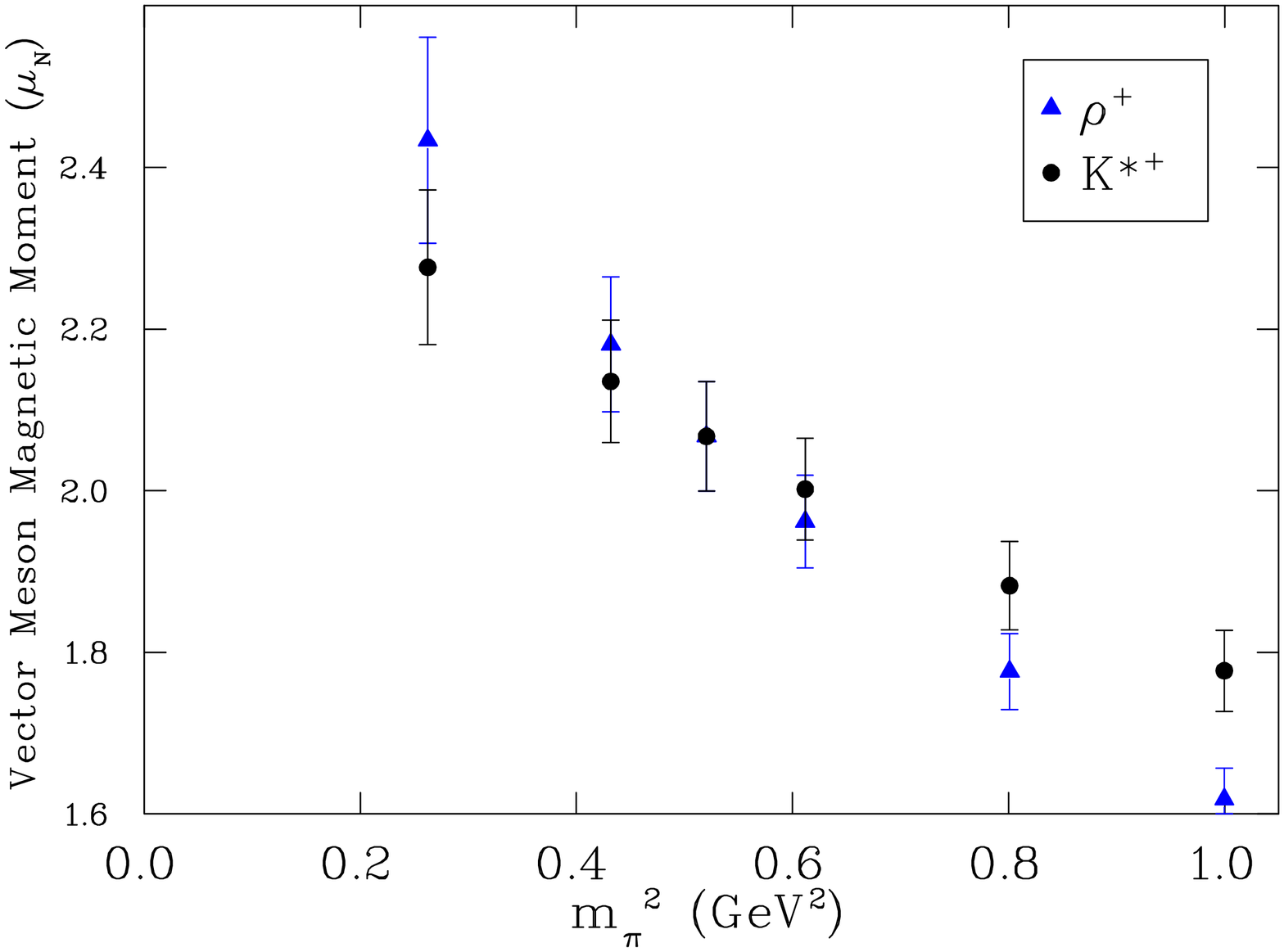,width=2.3in,angle=90}}
\caption{Magnetic moments (in nuclear magnetons) for $\rho^+$ and $K^{*+}$.}
\label{mag_vector}
\end{figure}

Note that the extracted g-factors are in the particle's natural 
magnetons.  To convert them into magnetic moments in terms of the commonly-used nuclear 
magnetons ($\mu_N$), we need to scale
the results by the factor $938/M$ where $M$ is the mass of the particle
measured in the same calculation at each pion mass.
Fig.~\ref{mag_vector} shows the results for ${\rho^+}$ and ${K^{*+}}$.
The different pion-mass dependence between ${\rho^+}$ and ${K^{*+}}$ mostly 
comes from that in the masses that are used to convert the g-factors to magnetic moments.
The values at the chiral limit extrapolated from Eq.~(\ref{g1}) are 
$\mu_{\rho^+}=3.25(3) \mu_N$ and $\mu_{K^{*+}}=2.81(1) \mu_N$. 
There is no experimental information on these quantities.
Compared to the form factor method (see Fig.7 in~\cite{Hedditch07}), 
our results are again a little higher.
At the strange pion mass point (the 3rd data point from the left), the two coincide to
give a prediction for the magnetic moment of the $\phi(1020)$ meson, $\mu_\phi=2.07(7) \mu_N$.

%
%
%
%

\begin{figure}[!htp]
\centerline{\psfig{file=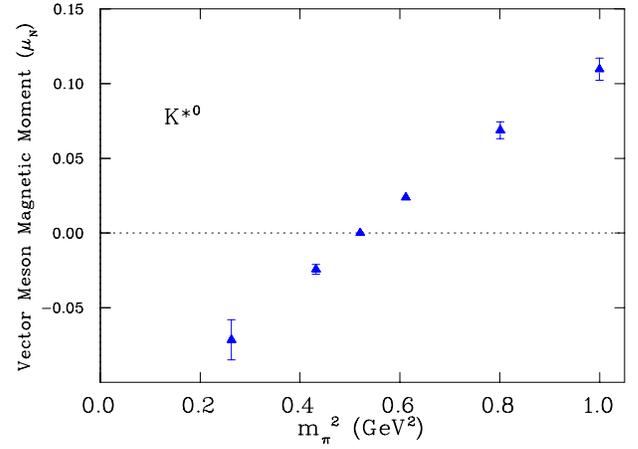,width=2.3in,angle=90}}
\caption{Magnetic moments (in nuclear magnetons) for $K^{*0}$.}
\label{mag_vector1}
\end{figure}

Fig.~\ref{mag_vector1} shows the results for ${K^{*0}}$.
Our results confirm the expectation that $\mu_{K^{*0}}$ is small but has an interesting 
quark mass dependence.
It is positive when the d-quark is heavier than the s-quark, 
exactly zero when they are equal, and turns negative when the d-quark is lighter than the s-quark.
The same behavior has been observed in the form factor method (see Fig.11 in~\cite{Hedditch07}).

%
%

\clearpage
 \subsection{Axial mesons}

\begin{figure}[!htp]
\parbox{.5\textwidth}{%
\psfig{file=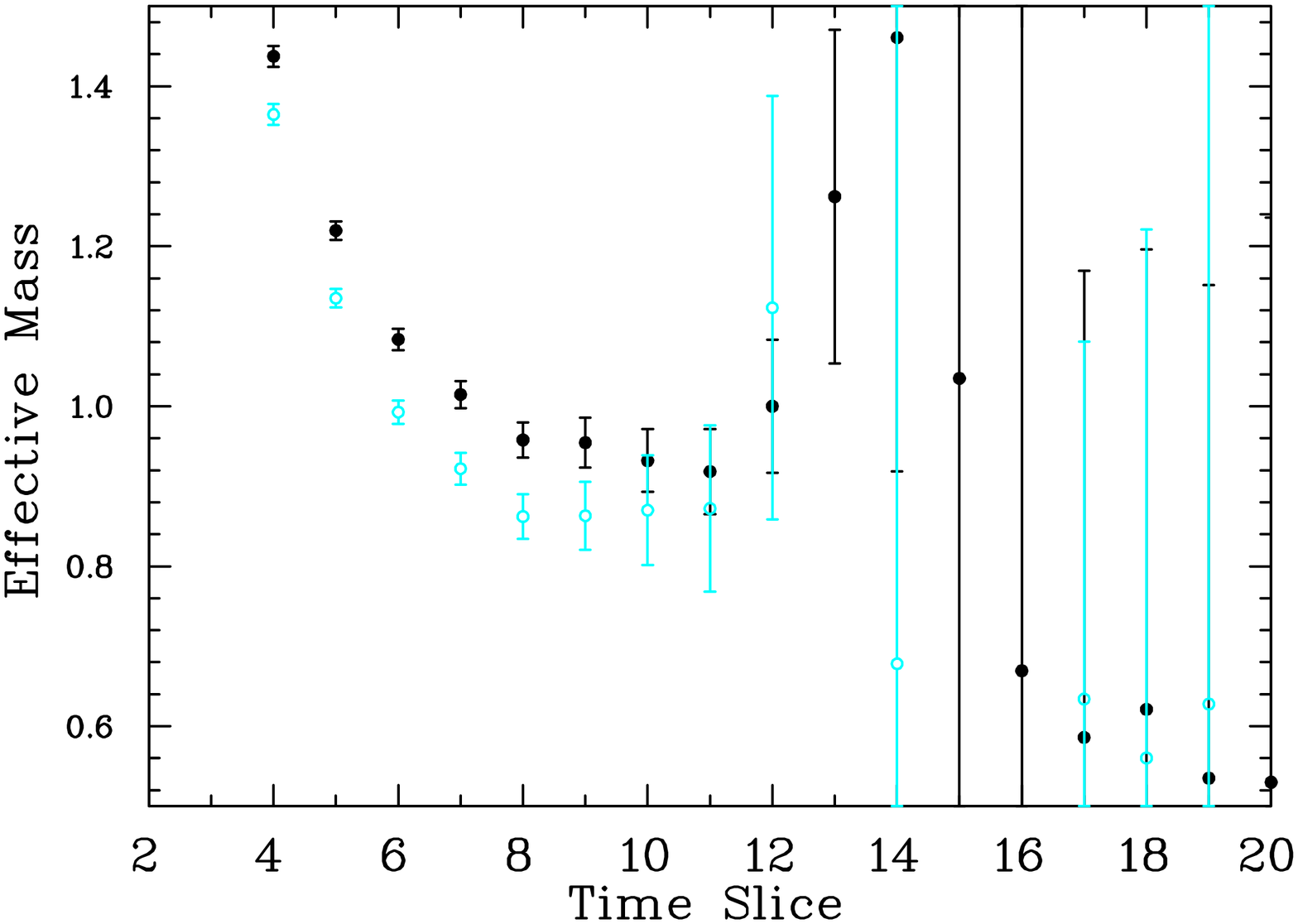,width=2.3in,angle=90}}
\parbox{.5\textwidth}{%
\psfig{file=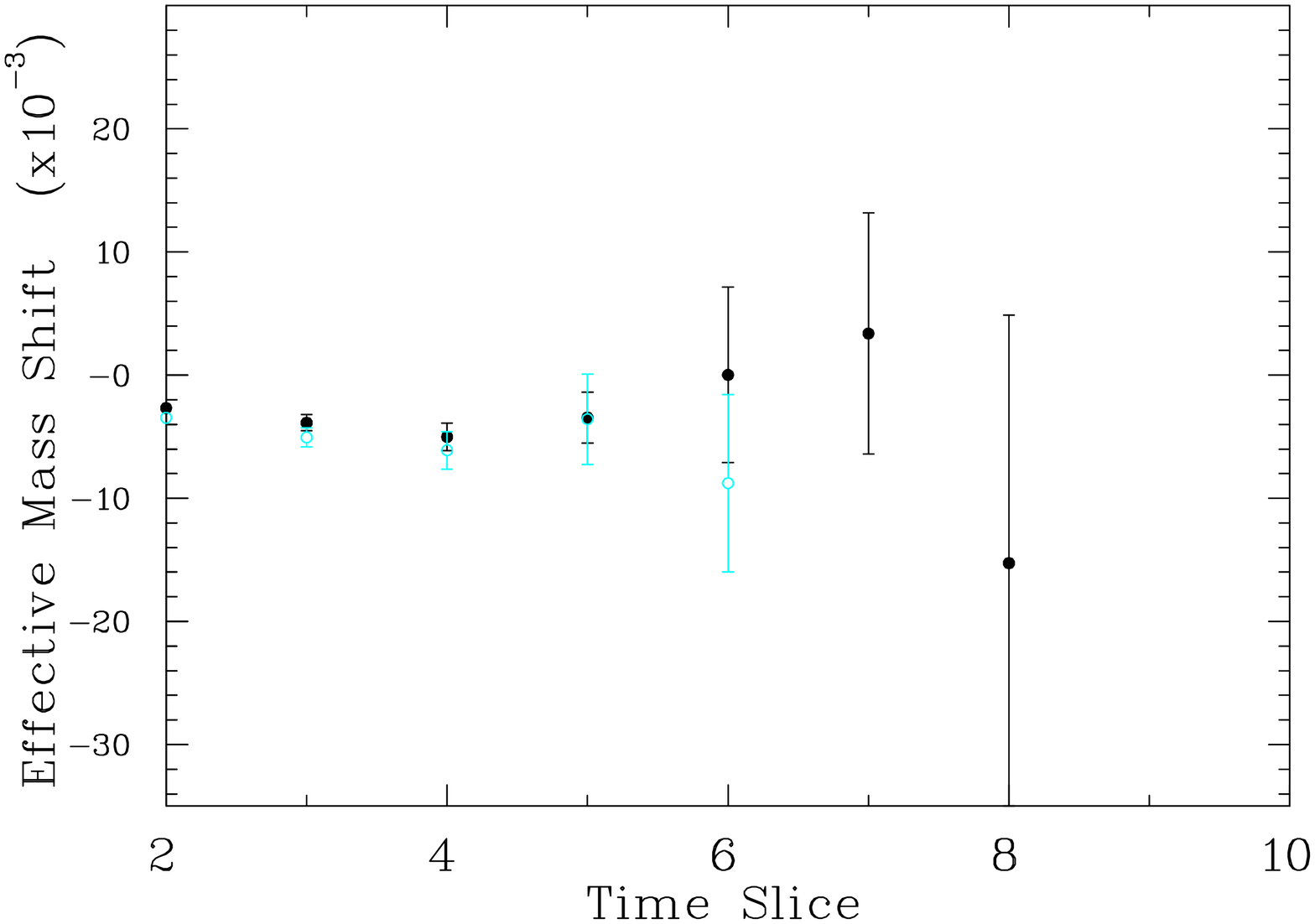,width=2.3in,angle=90}}
\caption{Effective mass plot for the $a_1^+$ axial meson mass
at zero field (top), and effective mass shifts at the 2nd weakest magnetic field (bottom)
in lattice units. The solid and empty symbols correspond to the heaviest 
and 2nd lightest pion masses, respectively.}
\label{emass2-ax}
\end{figure}

\begin{figure}[!htp]
\parbox{.5\textwidth}{%
\centerline{\psfig{file=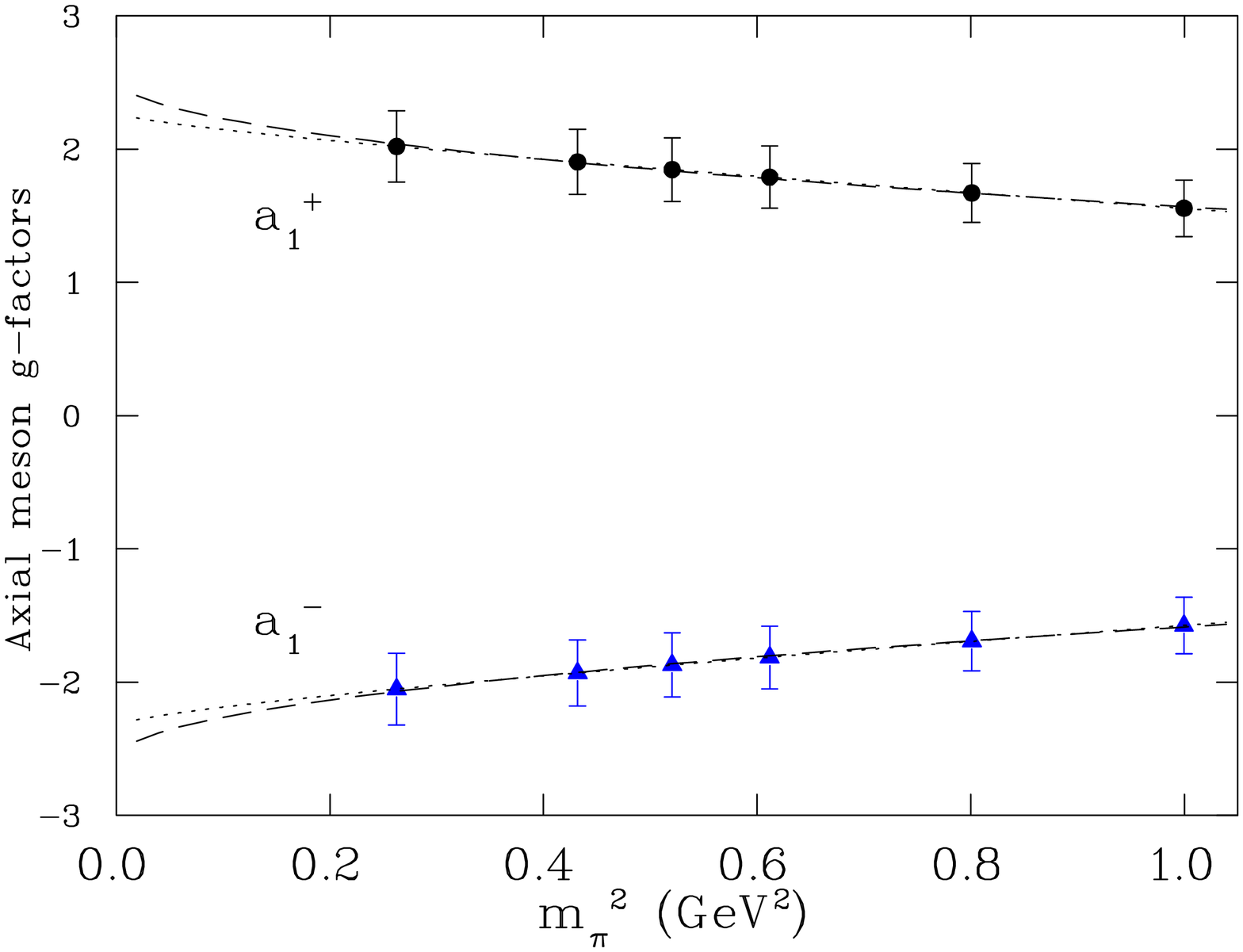,width=2.3in,angle=90}}}
\parbox{.5\textwidth}{%
\centerline{\psfig{file=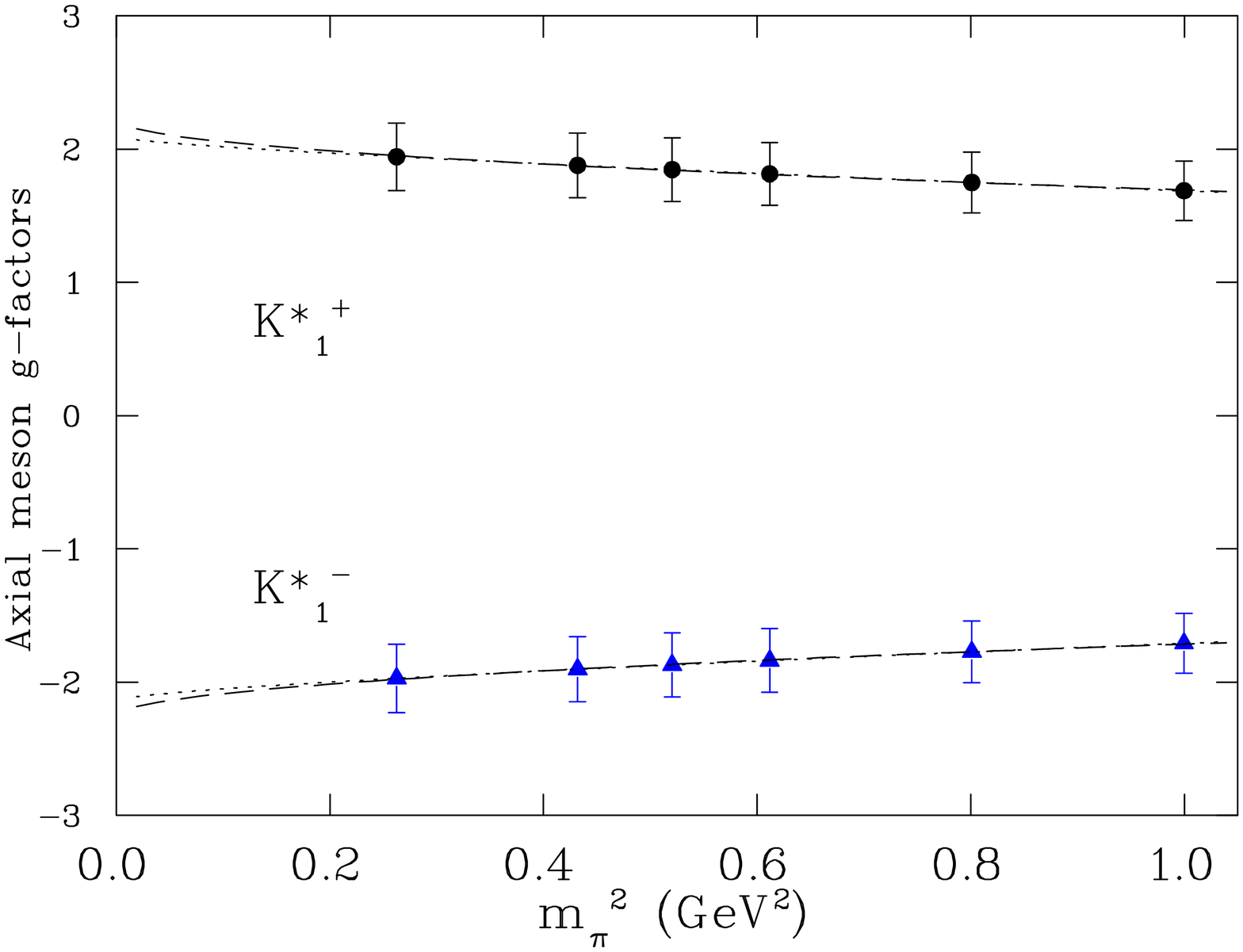,width=2.3in,angle=90}}}
\caption{G-factors for the $a_1^\pm$ (top) and $K_1^{*\pm}$ (bottom) axial mesons 
as a function of pion mass squared. 
The 2 lines are chiral fits according to
Eq.~(\protect\ref{chiral1}) (dashed), Eq.~(\protect\ref{chiral2}) (dotted).}
\label{gfac2}
\end{figure}

\begin{figure}[!htp]
\centerline{\psfig{file=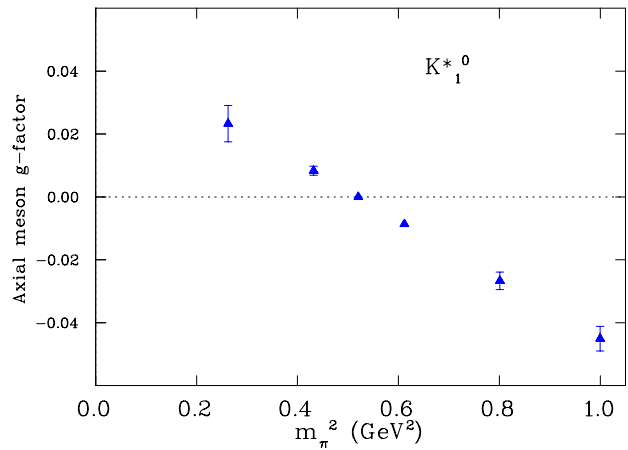,width=2.3in,angle=90}}
\caption{G-factor for the neutral axial meson $K_1^{*0}$.}
\label{gfac2n}
\end{figure}

Fig.~\ref{emass2-ax} shows the effective mass shifts for the 
$a_1^+$ axial meson at the 2nd value of the magnetic field ($eBa^2=0.00216$).
The signal is noisier compared to the vector case, 
but a plateau is still visible between time slice 
3 to 5 in the mass shifts.
Fig.~\ref{gfac2} shows the g-factors for $a_1^\pm$ and $K_1^{*\pm}$ 
extracted from this window at the six pion masses.
The g-factors are very similar to their counterparts in the vector channel (see Fig.~\ref{gfac_vector}).
Fig.~\ref{gfac2n} shows the g-factors for $K_1^{*0}$. They are small as expected, 
but have a linear behavior across the zero as a function of the pion mass squared. 
Interestingly, they have the 
opposite sign to that in the vector channel (see Fig.~\ref{mag_vector1}):
negative when the d-quark is heavier than the s-quark,
exactly zero when they are equal, and turns positive when the d-quark is lighter than the s-quark.

%
%

%
%

%
%

 \subsection{Tensor mesons}

\begin{figure}[!htp]
\parbox{.5\textwidth}{%
\psfig{file=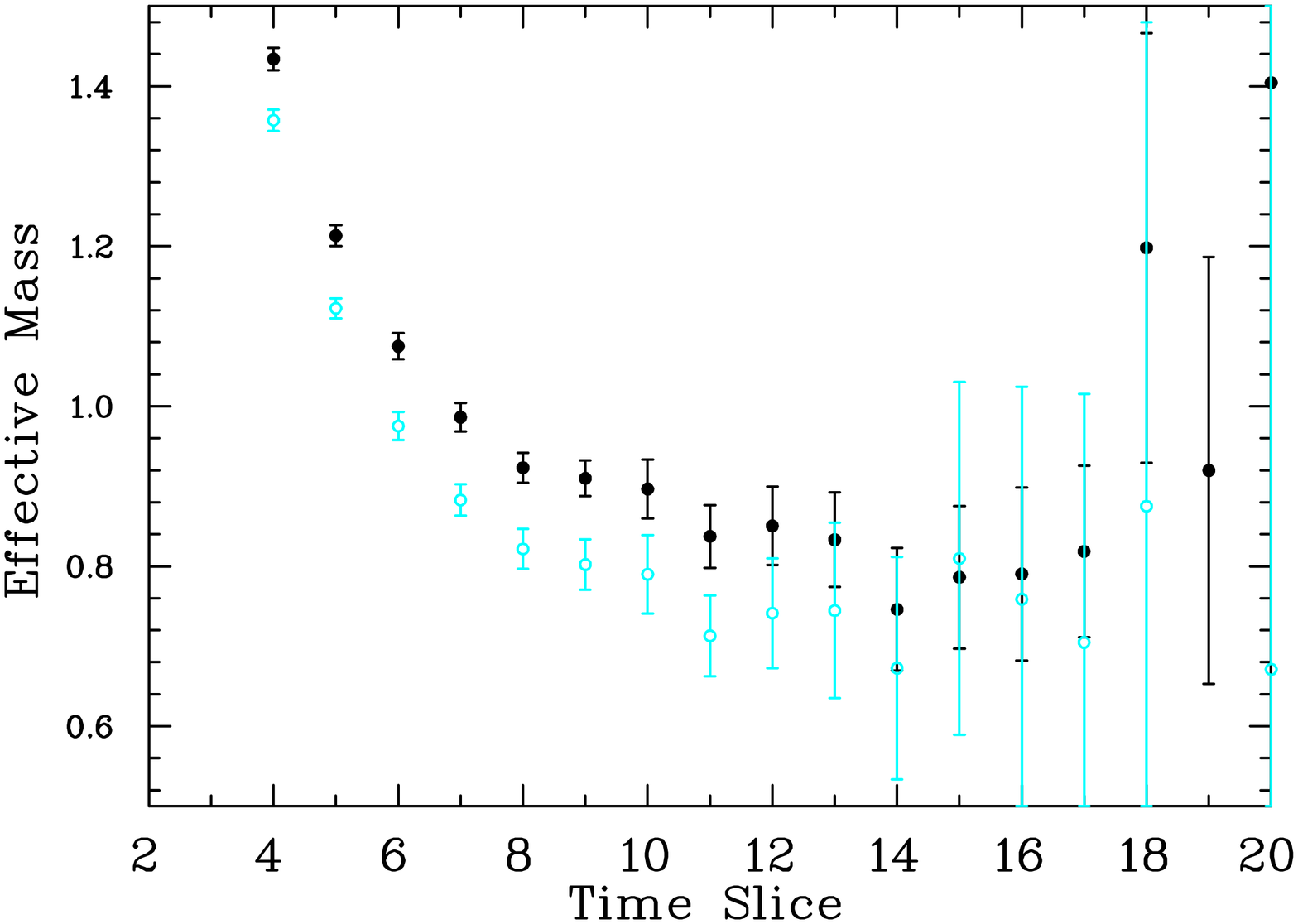,width=2.3in,angle=90}}
\parbox{.5\textwidth}{%
\psfig{file=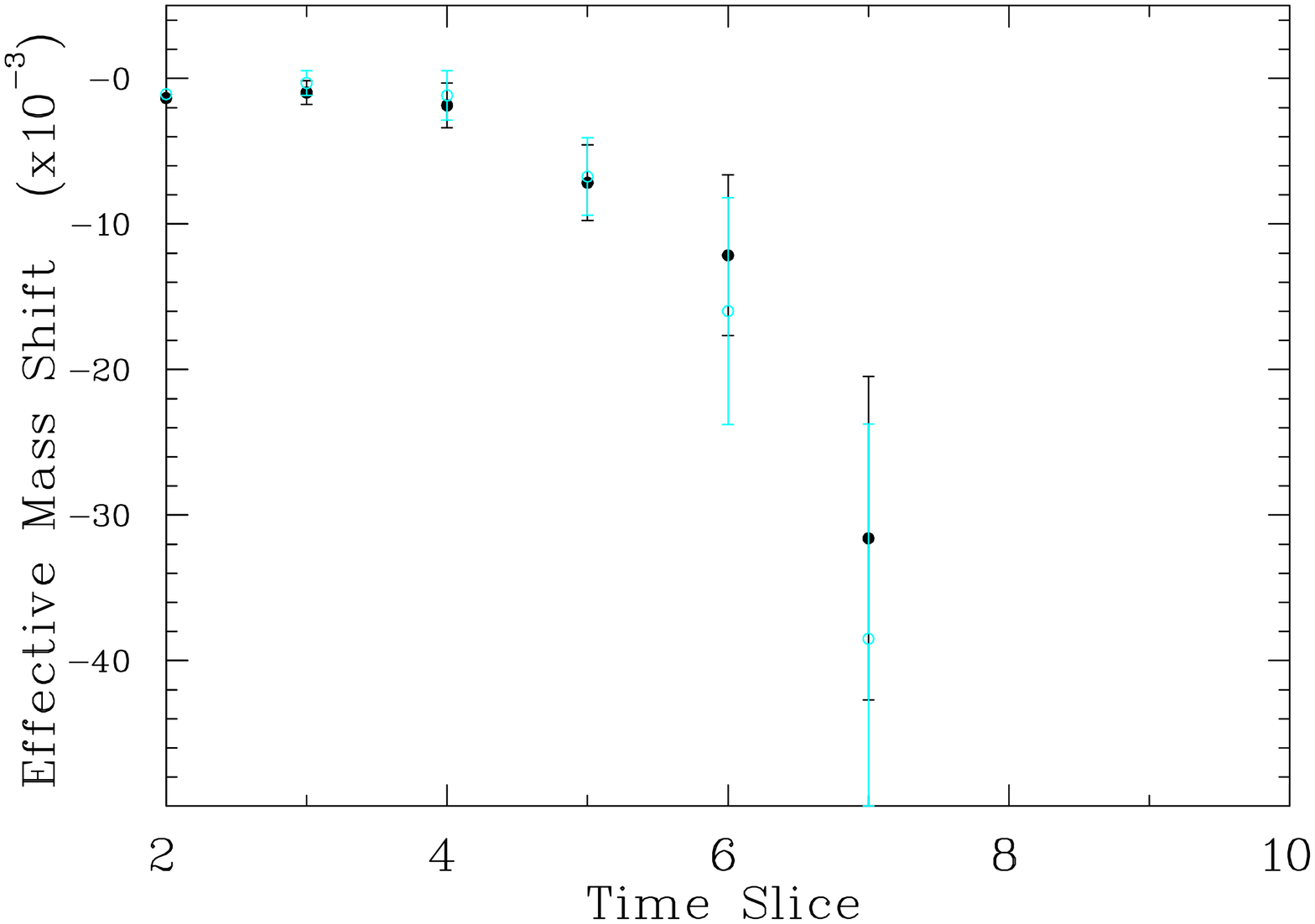,width=2.3in,angle=90}}
\caption{Effective mass plot for the $b_1^+$ tensor meson mass
at zero field (top), and effective mass shifts at the 2nd weakest magnetic field (bottom)
in lattice units. The solid and empty symbols correspond to the heaviest 
and 2nd lightest pion masses, respectively.}
\label{emass2-ts}
\end{figure}

\begin{figure}[!htp]
\parbox{.5\textwidth}{%
\centerline{\psfig{file=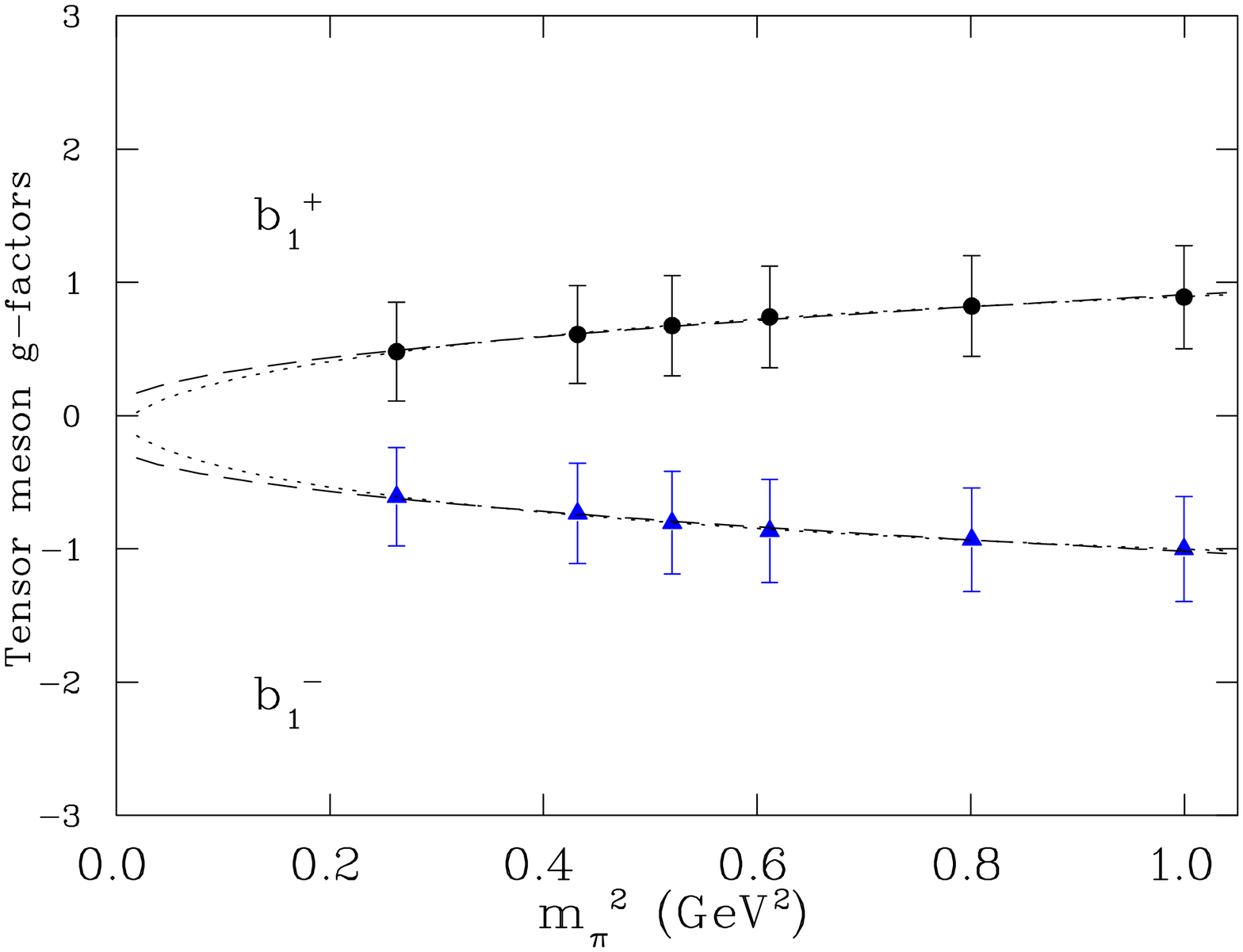,width=2.3in,angle=90}}}
\parbox{.5\textwidth}{%
\centerline{\psfig{file=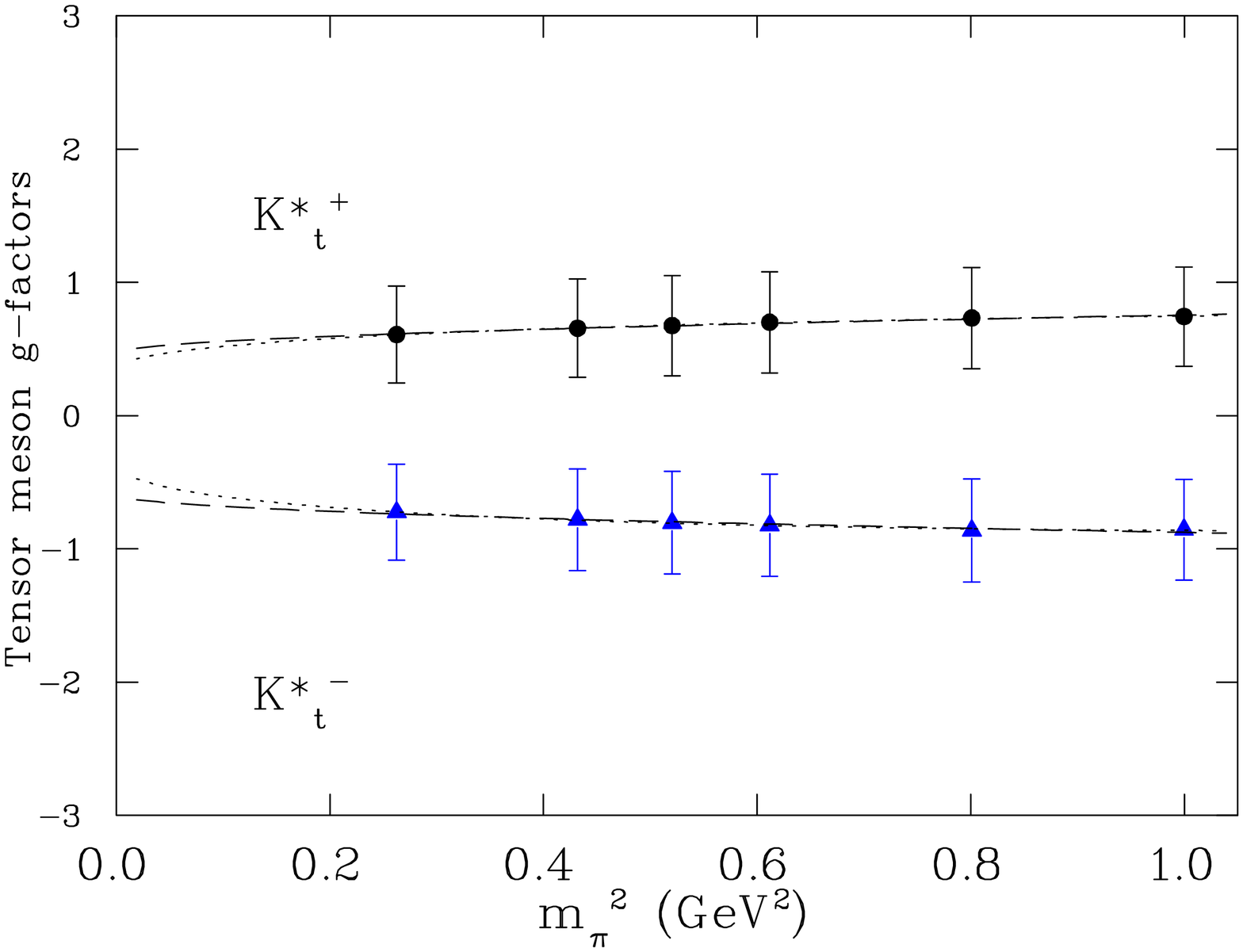,width=2.3in,angle=90}}}
\caption{G-factors for the charged tensor mesons $b_1^\pm$ (top) and $K_t^{*\pm}$ (bottom)
 as a function of pion mass squared. 
The 2 lines are chiral fits according to
Eq.~(\protect\ref{chiral1}) (dashed), Eq.~(\protect\ref{chiral2}) (dotted).}
\label{gfac3}
\end{figure}

\begin{figure}[!htp]
\centerline{\psfig{file=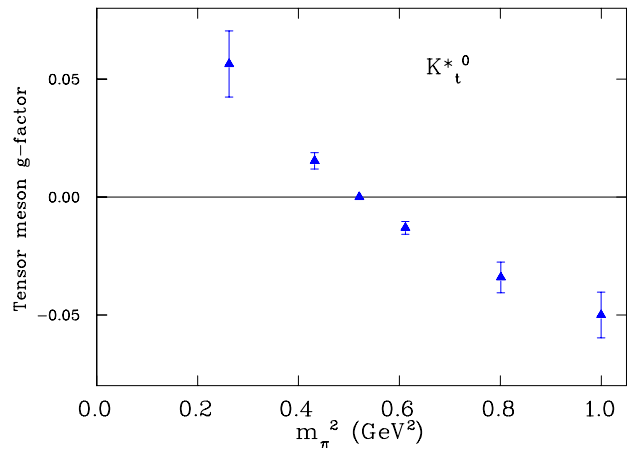,width=2.3in,angle=90}}
\caption{G-factor for the neutral tensor meson $K_t^{*0}$.}
\label{gfac3n}
\end{figure}

Fig.~\ref{emass2-ts} shows the effective mass shifts for the 
$b_1^+$ tensor meson at the 2nd value of the magnetic field ($eBa^2=0.00216$).
There is a signal, but much noisier than the axial case and  
there is barely a plateau in the mass shifts. If we fit the data 
between time slice 4 to 6, the results are shown in Fig.~\ref{gfac3} 
for $b_1^\pm$ and $K_t^{*\pm}$ 
extracted from this window at the six pion masses.
The g-factors have a weak pion mass dependence and relatively large errors. 
They have smaller values than the axial counterparts.
Fig.~\ref{gfac3n} shows the g-factors for the neutral $K_t^{*0}$, which 
displays a similar linear behavior as the axial counterpart.
This is evidence that there is indeed a signal in the tensor case.

%
%

%
%

%
%

\clearpage
 \section{Conclusion}
In conclusion, we have computed the magnetic moment of vector, axial and tensor 
 mesons
on the lattice using the background field method and standard lattice 
technology. Our results for the vector mesons are consistent with those 
from the form factor method where a comparison is possible.
The results for the axial and tensor mesons are new, although the latter still suffer 
from large errors.
Nonetheless, our results demonstrate that the method is robust and relatively 
inexpensive.  Only mass shifts are required. 
There is no experimental information on these quantities so the lattice 
results can serve as a guide from first principles.
Since the feasibility of the method is extended to the meson sector,
the calculation can be improved in a number of ways.
First, it should be repeated on a lattice of larger size 
in order to get an idea about finite-volume effects. More statistics are 
needed in the axial and tensor cases to better isolate the signals.
Second, there is a need to push the calculations to smaller pion masses 
so that reliable chiral extrapolations can be applied.
Third, the calculation should be extended to full QCD 
in order to see the effects of the quenched approximation, 
both in the SU(3) sector and in the U(1) sector. 
With the availability of dynamical configurations, all of the 
improvements can be made at the same time.
In particular, the U(1) effect in the sea quarks can be evaluated by re-weighting the 
determinants in the correlation functions, without the need to generate new 
dynamical ensembles.

\begin{acknowledgments}
This work is supported in part by U.S. Department of Energy
under grant DE-FG02-95ER40907.  W.W. acknowledges a research leave from Baylor University.
The computing resources at NERSC and JLab have been used.
\end{acknowledgments}

\end{document}